\def\be{\begin{equation}}
\def\ee{\end{equation}}
\def\ba{\begin{eqnarray}}
\def\ea{\end{eqnarray}}
\def\l{\label}
\def\n{\nonumber \\}
\def\b{\bibitem}
\def\EPC{{\em Eur. Phys.} C}
\def\PLB{{\em Phys. Lett.}  B}

\documentstyle[a4,12pt,epsfig,graphicx]{article}
 \begin{document} 

\title{Dijets with 2,1 or 0 rapidity gap: Factorization breaking at the 
Tevatron}

\author{A.Bialas \\ M.Smoluchowski Institute of Physics \\Jagellonian
University, Cracow\thanks{Address: Reymonta 4, 30-059 Krakow, Poland;
e-mail:bialas@th.if.uj.edu.pl.} 
\\\\ R.Peschanski \\
CEA/DSM/SPhT, Saclay\\ Unit\'e de recherche 
associ\'ee 
au CNRS\thanks{Address: 
F-91191 Gif-sur-Yvette Cedex, France;
	e-mail: pesch@spht.saclay.cea.fr.}
}

\maketitle

\begin{abstract} 
Central  production cross-sections of hard dijets with 2,1 or 0 rapidity gap at 
Tevatron are analyzed in terms of diffractive (``a la Good-Walker'') and non 
diffractive fluctuations of the incident particles. The observed large 
factorization breaking and the unexpected high value of the 2 to 1 gap 
cross-section ratio are explained in terms of  scattering with and between the 
incident particles.
\end{abstract}
 
{\bf 1.} Recent measurements of production of two large transverse
momentum jets in the central rapidity region at Fermilab energies
\cite{F,G} revealed a strong breaking of the Regge Factorization between the
``no-gap'', ``single-gap'' and ``two-gap'' cross-sections. If the presence of
the rapidity gap is interpreted as the colour singlet ({\it cf.} ``Pomeron'') 
exchange and its
absence as the octet exchange, Regge factorization appears to be broken if

\ba
 \frac{\Sigma_2}{\Sigma_1} \equiv R_{2/1} \  \ne \  \frac{\Sigma_1}{\Sigma_0} 
\equiv R_{1/0}\ , \l{1}
\ea
where $\Sigma_0$, $\Sigma_1$ and $\Sigma_2$ denote the 
cross-sections
for ``no-gap'', ``single-gap'' and ``two-gap'' events, respectively, and 
$R_{i/j} \equiv {\Sigma_i}/{\Sigma_j},$ by definition.

Experimentally \cite{F}, the ratio $R_{2/1}$ (resp. $R_{1/0}$) in (\ref{1}) is 
estimated in practice by the ratio of 
dijet event rates per unit $\xi_p$  (resp. $\xi_{\bar p}$), the fraction of 
momentum lost by the incident proton  (resp. antiproton), measured as a 
function of $x_{Bj}.$ Averaged over (small) $x_{Bj},$  the results are quoted 
to be 
$R_{2/1}\sim 0.8 \pm 0.26$ for the L.H.S of (\ref{1})  and $R_{1/0}\sim 0.15 
\pm 0.02$ for the R.H.S. \cite{F}. The factorization breaking $
{R_{1/0}}/{R_{2/1}}\sim 0.19 \pm 0.07$ \cite{F}  thus reaches a     factor 
as small as  1 over 5. 
In itself, the high value of $R_{2/1} ={\cal O} (1) $ is
also a question to be 
understood. Our aim is to give a theoretical estimate  of these ratios.

In the present note we argue that the origin of these  effects  is
basically the same  as that observed in measurements of 
diffractive (virtual) photon-induded and hadron-induced processes
\cite{HF,G}. We show that it can be described in  the same framework  as done 
in Ref. \cite{B} as due to the influence on dijet diffractive production of 
the soft 
scattering of the diffracted initial particles (protons and antiprotons at 
Tevatron).

{\bf 2.} Let us recall briefly the argument given in \cite{B} where the
(single) diffractive production of two large transverse momentum jets
was discussed. Following the idea of Good and Walker \cite{GW}, one
first expands the incident hadron state into a superposition of
diffractive (eigen)states. 
\ba
|h>= c_1|\psi_1> + c_2|\psi_2> +...  \l{2}
\ea

A diffractive state is  an eigenstate (by definition, in the subspace of 
states spanned by diffractive interactions) of absorption, i.e.
we have
\ba
T|\psi_i> =\lambda_i |\psi_i> + \ \omega_i |\phi_i>   \l{3}
\ea
\noindent where $T=1-S$ is the absorption operator and thus $\lambda_i$ is the
absorption coefficient\footnote{At high energy $\lambda_i$ is dominantly
real and positive.} of the state $|\psi_i>$. 
$|\phi_i>$ describes the
inelastic states which are at the origin of absorption. They are
multiparticle states which do not show any rapidity gap\footnote{
The first term in (\ref{3}) is usually interpreted as
``Pomeron exchange'' where no colour is exchanged between $|\psi_i>$ and
the target. The second term would then describe all colour exchanges and
thus $|\omega_i|^2$ represents the probability of this colour-exchange
interaction.}. 
Unitarity of the S-matrix implies 

\ba
|1-\lambda_i|^2 =1- |\omega_i|^2 =1-\sigma^i_{nondiff}\ ,
\l{4} 
\ea
where $\sigma_{nondiff}$  is the non-diffractive 
cross-section. (\ref{4}) is a 
special case of the generic relations for diffraction
\ba
|1-\lambda|^2 &=& 1-\sigma_{inel} = 1-\sigma_{nondiff}-\sigma_{dd}
\n 1-\omega^2 &=& 1-\sigma_{nondiff} = 1-\sigma_{inel}+\sigma_{dd}\l{4a} 
\ea
where $\sigma_{dd}$ is the diffractive dissociation cross-section.
Indeed, 
for a diffractive state, as seen from (\ref{3}), $\sigma_{dd}=0$.

Using (\ref{2}) and (\ref{3}), it is not difficult to express the
diffractive transitions between different hadronic states in terms of
the absorption ($\lambda_n$) and expansion ($c_n$)  coefficients:
\ba
<h'|{\bf t}|h> = \sum_n \lambda_n \ (c_n')^*c_n   \ .   \l{5}
\ea

The physical content of the expansion (\ref{2}) and of the formula
(\ref{5}) depends, of course, on the physical meaning one ascribes to
the ``diffractive'' states $|\psi_i>$. Following \cite{B} we could take them 
as
states with a fixed number and transverse positions (in impact parameter 
space) of partons
\cite{FV,MP}.
In a modern language, we could equivalently consider QCD dipole 
states \cite{BP}.

{\bf 3.}
We are interested in three\footnote{It will also be useful to  consider  the 
process {\it (ii)$^*$}  symmetric to {\it (ii)} in the interchange of left and 
right moving 
projectiles, namely \newline
\centerline{$(ii)^* \ \ p+\bar{p}\;\; \rightarrow 
\;\; p + (C) +\bar{p'}\  .$}}
 processes:
\ba
& &(i) \ \ \ p+\bar{p}\;\; \rightarrow \;\; p + (C) +\bar{p}  \n
& &(ii) \ \ p+\bar{p}\;\; \rightarrow \;\; p' + (C) +\bar{p}  \n
& &(iii)\  p+\bar{p}\;\; \rightarrow \;\; p' + (C) +\bar{p}' \ , \l{6}
\ea
where $(C)$ denotes a centrally produced system
(at rapidity $y$) of two large transverse momentum jets and a
``background'' of the soft particles nearby. $p'$, ($\bar{p}'$) represent
the states which  have no rapidity gap between
the $p$ ($\bar{p}$) and   $(C)$.
 Thus (i) corresponds to ``two-gap'' events, (ii) to
``single-gap'' events and (iii) to ``no-gap'' events, see Fig.1.

\begin{figure}[ht]
\centerline{
\epsfysize=5cm 
\epsfbox{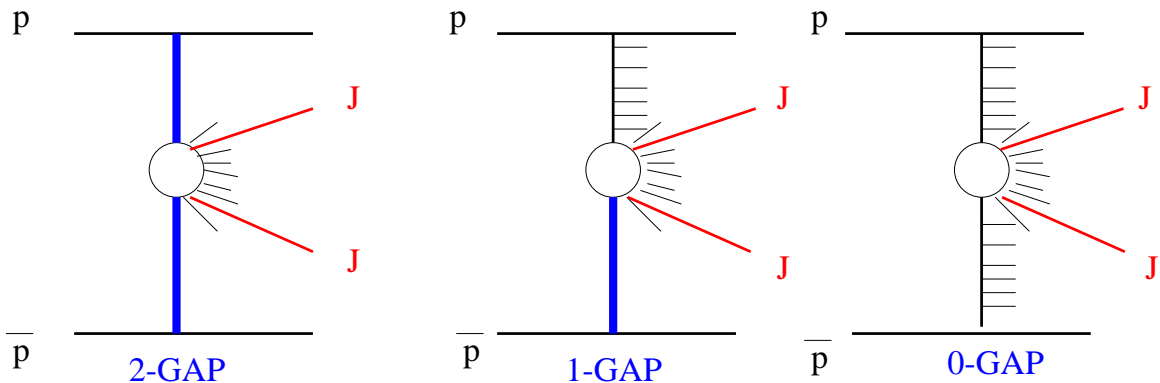}}
\caption{{\it Diffractive dijets with and without gaps.} }
\end{figure}

Now we have to write down the  expansions of (\ref{6}) into the
 diffractive states. Since we are interested in {\it hard}
diffraction, i.e., in the process whose probability is small, we follow
\cite{B} and assume that the expansion is quasi-diagonal.
Consequently we write
\eject
\ba
|p> &=& \ \ \ \ \ |g> +\  \epsilon |g + D> +\  \epsilon_P |p+ D> \n
|p'+C> &=& -\ \epsilon^* |g> +\ |g+D> +\ \epsilon' |p+D> \n
|p+C> &=& -\ \epsilon_P^* |g> -\ \epsilon'^*|g+D> +\  |p+D>    \l{7}
\ea
\noindent
where $\epsilon, \epsilon_P$ and $\epsilon'$ are small (and thus will 
be
kept only up to first order)\footnote{The relations between expansion
coefficients follow from orthonormality of the states.}. Here $|g>$
denote a superposition of diffractive states representing
 a bunch of {\it soft} partons (close to their
distribution in the  left moving proton) and $(D)$ is a superposition of
 partonic states consisting of
a {\it hard} ({\it i.e.} small in transverse space) dipole at rapidity $y$
and a number of
soft partons with rapidities nearby (this system
eventually decays predominantly 
into two large transverse momentum jets and the
background of soft particles).

Using (\ref{3}) and (\ref{7}) we thus obtain

\ba
<p|{\bf t}|p>&\equiv& \lambda_p=  \lambda_g\n
<p'+C|{\bf t}|p>&=&
\epsilon\ (\lambda_{(g+D)}-\lambda_g) \n
<p+C|{\bf t}|p>
&=&\epsilon_P\ (\lambda_{(p+D)}-\lambda_g), \l{8} 
\ea 
where the   transition matrix ${\bf t}$ is to be considered  diagonal in 
both  rapidity and impact parameter space. All quantities on the R.H.S.  
are to be understood as linear combinations of the
corresponding quantities defined for truely diffractive parton states. 

Following \cite{B} we now assume that $\lambda_{(g+D)}$ and
$\lambda_{(p+D)}$ {\it for diffractive states} 
can be calculated assuming an independent scattering of
their components \cite{GL}, i.e.
\ba
1-\lambda_{(g+D)}= (1-\lambda_{g})(1-\lambda_{D});\;\;\;
1-\lambda_{(p+D)}= (1-\lambda_{p})(1-\lambda_{D}).  \l{9b}
\ea
When this is inserted into (\ref{8}) we have
\ba
& &<p'+C|{\bf t}|p>=
\epsilon\ \lambda_D(1-\lambda_g) \n
& &<p+C|{\bf t}|p>
=\epsilon_P\ \lambda_D(1-\lambda_g)\ . \l{8c} 
\ea 

{\bf 4.} To obtain the  cross-sections $\Sigma_{1,2}$ for one-gap and two-gap 
events,
 see (\ref{1}),
 we have to square the corresponding amplitudes, sum over the final
hadronic states and integrate over the suitable rapidity and impact parameter 
intervals. In a first stage we shall discuss the differential cross sections 
$\sigma_{1,2},$ before integration over rapidity and impact parameter 
variables. For given rapidity $y$ and impact parameter position 
$\vec{s},$ the summation extends for $\sigma_1$ over all states 
$(p')$ and $(C)$  as the detailed final state is not measured. For
$\sigma_2$ one sums only over all states of (C), because the final
proton is identified. These differences imply different averaging
procedures for the diffractive states. Taking this into account, we
obtain

\ba
\sigma_1(\vec{s},y)&=&\left[|\epsilon(\vec{s},y)|^2
|\lambda_{D}(\vec{b}\!-\!\vec{s},Y\!-\!y)|^2\right]_{av}
\left[|1-\lambda_g(\vec{b},Y)|^2\right]_{av} ;\n
\sigma_2(\vec{s},y)&=&\left[|\epsilon_P(\vec{s},y)|^2
|\lambda_{D}(\vec{b}\!-\!\vec{s},Y\!-\!y)|^2\right]_{av}
|1-\left[\lambda_g(\vec{b},Y)\right]_{av}|^2\ ,  
 \l{9c}
\ea
where we have explicitely indicated the averaging procedures. $Y,\vec{b}$ are 
the rapidity and impact parameter distances between the incident particles. At 
this point
one can  observe that, as seen from (\ref{4}), (\ref{4a}), 
\ba
\left[|1\!-\!\lambda_g(\vec{b},Y)|^2\right]_{av}
\!\!=1\!-\!\left[\omega^2_g(\vec{b},Y)\right]_{av}\!\!=1\!-\!\omega^2_p 
(\vec{b},Y) =
\left(1\!-\!\left[\lambda_g(\vec{b},Y)\right]_{av}\right)^2\!\! + \sigma_{dd}. 
\l{8d}
\ea
It is important to remember that - since the averaging concerns only 
one vertex -  $\sigma_{dd}$ here corresponds to  the {\it single 
diffractive dissociation in one vertex}. All this implies that 
\ba
\sigma_1(\vec{s},y)=\left[|\epsilon(\vec{s},y)|^2|
|\lambda_{D}(\vec{b}\!-\!\vec{s},Y\!-\!y)|^2\right]_{av}
[1-\omega^2_p (\vec{b},Y)]   \l{9x}
\ea
so that in the following one can omit the index ${av}$ without running
into confusion.

One sees that the formula (\ref{9c}) for $\sigma_2$ is
not symmetric with respect to the interchange of the projectile
and the target. To restore the symmetry we have to require that
\ba
|\lambda_{D}(\vec{b}\!-\!\vec{s},Y\!-\!y)|^2= V_P
\ |\epsilon_P(\vec{b}\!-\!\vec{s},Y\!-\!y)|^2    \l{99}
\ea
where, {\it a priori}, $V_P$ is a 
vertex function of $(\vec{s},y;\vec{b}\!-\!\vec{s},Y\!-\!y)$ which must be 
symmetric with respect to exchange of the
corresponding arguments. However, since the L.H.S. of (\ref{99}) depends
only on $\vec{b}\!-\!\vec{s}$ and $Y\!-\!y$, $V_P$ can depend neither
on $\vec{s}$ nor on $y$. Symmetry implies that it must be a constant,
depending only on the internal variables of the vertex.

The formula (\ref{99}) shows that the two-gap diffractive
interaction can be equivalently understood as the elastic 
 interaction of the projectile  with a
Good-Walker (Pomeron-like) fluctuation of the target.

From (\ref{9c}) and (\ref{99}) we deduce that
\ba
\sigma_2(\vec{s},y)&=&V_P\ \ \ |\epsilon_P(\vec{s},y)|^2
\ |\epsilon_P(\vec{b}\!-\!\vec{s},Y\!-\!y)|^2|1-\lambda_g(\vec{b},Y)|^2 \n
 &=&
V_P^{-1}\ |\lambda_D(\vec{s},y)|^2
\ 
|\lambda_D(\vec{b}\!-\!\vec{s},Y\!-\!y)|^2|1-\lambda_g(\vec{b},Y)|^2\ .
 \l{9i}
\ea

{\bf 5.} We have to discuss now the no-gap events and thus to go beyong
the Good-Walker argument, as the latter refers only to diffractive
interactions. To this end we observe that the probability of a no-gap
event to occur can be written as the product of (a) the probability of
fluctuation of the projectile into $|g'+D>$, i.e. $|\epsilon|^2$, and
(b) the probability of {\it non-diffractive} interaction of the central 
system $(D)$  with the target, i.e. $\omega^2_{D}$:
\ba
\sigma_0=  |\epsilon(\vec{s},y)|^2\ \omega^2_{D}(\vec{b}\!-\!\vec{s},Y\!-\!y)  
\l{15}
\ea
Since the same argument can also be used by exchanging the roles of the
target and projectile we also have
\ba
\sigma_0=  |\epsilon(\vec{b}\!-\!\vec{s},Y\!-\!y)|^2\ \omega^2_{D}(\vec{s},y)  
\l{16}
\ea
which implies that
\ba
[\omega_{D}(\vec{s},y)]^2 = V \ 
|\epsilon(\vec{s},y)|^2 
  \l{17}
\ea
where $V$ is another vertex  function,
 symmetric with respect to exchange of  the corresponding arguments. 
For the similar reasons as those following Eq.(\ref{99}) we deduce
that it is actually  a constant. Thus we finally obtain
\ba
\sigma_0&=&  V\ \ \ 
|\epsilon(\vec{s},y)|^2\ |\epsilon(\vec{b}\!-\!\vec{s},Y\!-\!y)|^2 \n
&=&
V^{-1}\ \omega_D^2(\vec{s},y)\ \omega_D^2(\vec{b}\!-\!\vec{s},Y\!-\!y)\ .
   \l{18}
\ea

 A closer look shows that the vertex functions $V$ and $V_P$
are identical. This can be seen by observing that the cross-section for 
one-gap events, $\sigma_1$, given in (\ref{9c}), can be also calculated 
as a product of 
(a) the probability of fluctuation of the target into $(\bar{p}+D)$, i.e. 
$|\epsilon_P|^2$, (b) the probability of {\it non-diffractive} interaction 
of $(D)$  with the projectile, i.e. $\omega^2_{D}$ and (c) the probability
that no {\it non-diffractive} interaction of the final antiproton
 with the projectile took place\footnote{Diffractive excitation of the 
projectile  is allowed.}, i.e. 
$(1-\omega^2_{\bar{p}})$ (c.f. (\ref{4a})). Thus we can  write
($\omega_{\bar{p}}= \omega_{{p}}$)
\ba
\sigma_1(\vec{s},y)
 =\ \left|\epsilon_P(\vec{b}\!-\!\vec{s},Y\!-\!y)\right|^2 
\ \omega^2_{D}(\vec{s},y)
\left[1-\omega^2_p(Y)\right] \ .  \l{9d}
\ea

Comparing (\ref{9d}) and (\ref{9x})  we have
\ba
|\epsilon_P(\vec{b}\!-\!\vec{s},Y\!-\!y)|^2\ 
[\omega_{D}(\vec{s},y)]^2=|\epsilon(\vec{s},y)|^2
\ 
[\lambda_D(\vec{b}\!-\!\vec{s},Y\!-\!y)]^2 \ . \l{14}
\ea

This equation has a simple physical meaning. It says that the
probability of a fluctuation of the projectile into the
projectile+gap+D configuration is proportional to the elastic
interaction in the projectile-D system, whereas the probability of
fluctuation into no-gap configuration is proportional to the
non-diffractive interaction in this system. It thus provides the
interpretation, in the Good-Walker language, of the ``Pomeron exchange''
mechanism for gap creation.

From (\ref{99}), (\ref{17}) and (\ref{14}) one sees that indeed
the vertex functions $V$ and $V_P$ are identical:
\ba
 V= V_P \ .\l{9g}
\ea
Using now (\ref{14}) and (\ref{9d}) we obtain\footnote{We can equivalently 
consider the symmetric $\sigma_1^*$ of $\sigma_1$ by interchange of left and 
right moving particles, namely
$$\sigma_1^*(\vec{s},y)\equiv \sigma_1(\vec{b}\!-\!\vec{s},Y\!-\!y)\ .$$}
\ba
\sigma_1(\vec{s},y)&=& V
\ \ \  
|\epsilon(\vec{s},y)|^2\epsilon_P(\vec{b}\!-\!\vec{s},Y\!-\!y)|^2
[1-\omega_p(\vec{b},Y)^2] \n
&=& 
V^{-1}\ 
|\omega_D(\vec{s},y)|^2|\lambda_D(\vec{b}\!-\!\vec{s},Y\!-\!y)|^2 
[1-\omega_p(\vec{b},Y)^2] \ .
 \l{9h}
\ea

{\bf 6.} Putting together the formulae  Eqs. (\ref{9i}),(\ref{18}) and 
(\ref{9h}) for the differential cross-sections, and even before evaluating 
directly the observable quantities  mentionned in section {\bf 1}, it is already 
possible to  get important qualitative hints on the pattern of factorization 
breaking in diffractive dijet production at the Tevatron.

Considering for instance, the differential\footnote{Not to be confused with the 
integrated ones (\ref{1}).} ratios $r_{i/j}\equiv 
\sigma_i/\sigma_j,$ we find
\ba 
r_{2/1} &=& \frac {|\epsilon_P(\vec{s},y)|^2}{|\epsilon(\vec{s},y)|^2|}\ \frac 
{|1-\lambda_g(\vec{b},Y)|^2}{\left[1-\omega^2_p(Y)\right]} = \frac 
{|\lambda_D(\vec{s},y)|^2}{|\omega_D(\vec{s},y)|^2}\ \frac 
{|1-\lambda_g(\vec{b},Y)|^2}{\left[1-\omega^2_p(Y)\right]} \n
r_{1}*_{/0} &=& \frac {|\epsilon_P(\vec{s},y)|^2}{|\epsilon(\vec{s},y)|^2}\  
{\left[1-\omega^2_p(Y)\right]} = \frac 
{|\lambda_D(\vec{s},y)|^2}{|\omega_D(\vec{s},y)|^2}\  
{\left[1-\omega^2_p(Y)\right]} \ ,
\l{ratios}
\ea
where we make use of $\sigma_1^*,$  ({\it cf.} footnotes $3,6$) instead of 
$\sigma_1$ in 
the second row for formal convenience,
the qualitative conclusions being unchanged.

Formulae (\ref{ratios}) show explicitely that the
factorization formula (\ref{1}) is violated:
\ba
\frac {\sigma_1^*(\vec{s},y)}{\sigma_0(\vec{s},y)}=
\frac {\sigma_2(\vec{s},y)}{\sigma_1(\vec{s},y)}\times 
\frac {[1-\sigma_{nondiff}(\vec{b},Y)]^2}{1-\sigma_{inel}(\vec{b},Y)}\ . 
\l{6.1}
\ea


We have expressed the coefficients $\lambda _g,\omega _p$ in terms of their 
physical interpretation (\ref{4}),(\ref{5}).
One sees that the factorization 
violating factor has the same origin as
that which was shown in \cite{B} to be reponsible for violation of
factorization between single diffractive processes at HERA and at
FERMILAB. The numerical value may be, however, somewhat different,
depending on the size of the cross-section for the {\it soft}
diffractive dissociation in $p\bar{p}$ collisions (giving the 
difference\footnote{In \cite{B}
this difference was neglected. If taken into account, the corresponding
factor between the HERA and FERMILAB cross-sections should be
$1-\sigma_{nondiff}$ rather than $1-\sigma_{inel}$. At small impact
parameters, however, which are most important for numerical estimates,
this difference is expected to be  small.}
between $\sigma_{inel}$ and $\sigma_{nondiff}$).

Another interesting remark individually concerns the  ratios $r_{2/1}$ and 
$r_{1}*_{/0}.$ Rewriting  expression (\ref{ratios}) as
\ba 
r_{2/1} &=& \frac {|\epsilon_P|^2}{|\epsilon|^2}(\vec{s},y)\   \frac 
{1-\sigma_{inel}}{1-\sigma_{nondiff}}  \n
r_{1}*_{/0} &=& \frac {|\epsilon_P|^2}{|\epsilon|^2}(\vec{s},y)\  
(1-\sigma_{nondiff}) \ .
\l{ratiosbis}
\ea
In fact, we have to take into account that the specific factor  
${|\epsilon_P|^2}/{|\epsilon|^2}$ is factorizable since it reflects the 
probability ratio for  a color singlet over an octet to be coupled to the 
projectile. Hence, 
we 
see that the essential factorization breaking factor is concentrated in the 
second ratio $r_{1}*_{/0}$ and it has the   same content as the corresponding
factor between the HERA and FERMILAB cross-sections \cite{B}. On contrary, the 
 factorization breaking factor in $r_{2/1}$ is mild, since it depends only on 
the difference between $1-\sigma_{inel}$ and $1-\sigma_{nondiff}$ due to  the 
diffraction dissociation contribution in the total cross-section. These 
conclusions will be confirmed by the following phenomenological application.

{\bf 7.} To obtain numerical estimates of the cross-sections $\Sigma_{i}$ and 
of the measured  ratios $R_{i/j},$ {\it cf.} section {\bf 1}, one has to
perform integration over the rapidity interval $y$ and   impact parameters 
$\vec{b}$ and $\vec{s}$.

For rapidity intervals, the integration is, in some sense, taken into account 
by the measurement itself, since event densities  by unit  of $\xi$ 
are given. hence, even if a full simulation would be welcome to match precisely 
the experimental conditions, the rapidity dependence is considered to be 
already integrated out. 

The impact parameter dependence has to be taken into account. This is difficult 
because several necessary elements are not known, so
that -at best- only a rough estimate is possible. Only the elastic $p\bar{p}$
amplitude is known with some precision:
\ba
\lambda_p(\vec{b},Y) = a_0 e^{-b^2/2B_{el}}
;\;\;\; a_0= \frac{\sigma_{tot}}{4\pi B_{el}}\ . \l{6.2}
\ea
At $\sqrt{s}=1800$ GeV, $\sigma_{tot}=71.71 \pm 2.02\ mb$  and         
$B_{el}=16.3 \pm .5  \  $GeV$^{-2}$ \cite{OR},
 so that $a_0 \approx$ 0.85. There is also the inequality (Pumplin
bound \cite{pum})
\ba
\lambda_D^2\leq \omega_D^2 \ .  \l{6.2a}
\ea
To simplify the discussion we assume tentatively
 that the shapes (in impact parameter space) of $\lambda_D^2$ and
$\omega_D^2$ are  similar and take the Gaussian forms for easy integration:
\ba
\left[\lambda_D(\vec{s},y)\right]^2=\eta \ \omega_D^2
(\vec{s},y)= A \exp[-s^2/B_D].  \l{6.3}
\ea
where $B_D$ is the slope in the elastic scattering of the central system
$(D)$. Guided by the results from hard diffraction of  virtual
photons \cite{phot}, we estimate $B_D$ to be in the region around
 4 GeV$^{-2}$ .

Using (\ref{9i}), this allows to determine $\sigma_2$. The integration over 
 $d^2s$ and $d^2b$ gives
\ba
\frac{\sigma_2}{\sigma_0}\equiv  R_{2/0} = \eta^2
\left(1-2\frac{a_0}{1+\zeta}+\frac{a_0^2}{1+2\zeta}\right)  \l{6.3a}
\ea
where $\zeta= B_D/B_{el}.$ 
 Using the obvious identity
\ba
 R_{2/1}= 
 R_{2/0}R_{0/1}= \frac{R_{2/0}}{R_{1/0} }     \l{6.3b}
\ea
one can thus calculate $R_{2/1}$ from (\ref{6.3a}) and the measured
ratio $R_{1/0}$.

The Fermilab measurements \cite{F} give $R_{1/0}\approx 0.15 \pm 0.02$.
Substituting all this into (\ref{6.3b}) 
  and  using the measured ratio $R_{2/1}= 0.8\pm
0.26$ we deduce that for $\zeta \approx  0.25$, $\eta \approx 1$
 (with an error of about  30\%). Although this value of $\eta$ is a  
perfectly acceptable one\footnote{The fact that we can take $\eta \sim {\cal 
O}(1/2-1)$  is related to the fact
that the central system D contains some soft partons apart from the hard
dipole, {\it e.g. inclusive} diffractive production \cite{BPR}. Therefore we 
would predict that in an experiment which
will measure the "elastic" two jets {\it e.g. exclusive} diffractive production 
without accompanying soft hadronic radiation \cite{KMR},
one shall expect $\eta \ll 1$ and  falling with increasing transverse momentum 
of
the jets.}, it should be noted that it saturates  the unitarity limit. 
More precise data are needed, however, to consider this interesting property as 
really
established.

This result shows two points. First that our analysis can naturally
explain a rather large value of the ratio $R_{2/1}$ measured by the
CDF coll. This seems to be a rather non-trivial result. As a sort of
by-product we obtain the second point, namely the strong violation of
the Regge factorization.

To obtain the theoretical value of the ratio 
\ba
R\equiv  R_{2/1}/R_{1/0} = R_{2/0}/R_{1/0}^2 ,\l{6.4x}
 \ea
  one needs additional
information about the soft single diffractive dissociation 
 cross-section in one vertex (c.f. Eq. (\ref{4})). 
For a rough estimate we take 

\ba
<\lambda_p^2(\vec{b})>= a_0^2\ e^{-b^2/ B_{diff}} \l{6.4} 
\ea 
which is a simplified version of the analysis by Miettinen and Pumplin
 \cite{MP}. From this and (\ref{6.2}) we deduce the diffraction dissociation 
cross-section 
 to be
 \ba
 \sigma_{dd}= (\kappa -1) \sigma _{el}     \l{6.4a}
 \ea
 where\footnote{The ratio $\kappa > 1$ {\it does not imply} that the slope in 
diffractive dissociation must be larger  that in elastic scattering.
This was 
explained already in Ref.[7]. The parameter
$B_{diff}$ is responsible for the 
behaviour of the average of sum of the
squares of the diffractive amplitudes 
(including elastic amplitude) {\it in impact parameter space}. From this 
information it is not possible to deduce directly the behaviour of diffractive 
dissociation channels in
momentum transfer. As shown by Ref.[7], although the 
diffractive dissociation {\it is} more preripheral in $b$ space than
elastic 
scattering, its slope in momentum transfer is smaller than that
of elastic 
scattering. As explicitely derived and clarified in the parton picture of 
Ref.[7], the slope in 
momentum transfer depends strongly on degrees of freedom which are other than 
the transverse ones and thus {\it is not given} by the fourier transform of 
(\ref{6.4}) ({\it cf.}  the comparison of  Fig. 2 (for 
$d\sigma/d^2\vec b$) and Fig. 5  (for $d\sigma/dt$) in Ref.[7]).} $\kappa= 
B_{diff}/B_{el}$. 
 This gives 
\ba
R_{1/0}=\eta\left( 1-2\frac{a_0}{1+\zeta}+ \frac{a_0^2}{1+2\zeta/
\kappa} \right)  \l{6.5}
\ea
One sees that the ratio  $R$ is independent of $\eta$ and thus completely 
determined by the value of $\kappa$. To estimate $\kappa$ we observe 
that the total {\it soft}
diffractive dissociation cross-section
 is approximately equal to the elastic one. Assuming the 
 approximate factorization between the single-diffractive and 
 double-diffractive cross-sections we can write for the sum of all contributions 
to the total  {\it soft} diffractive cross-section $$\sigma_{difftot}/\sigma 
_{el} = 2(\kappa -1) + (\kappa -1)^2  =(\kappa^2-1) \approx 1$$ leading to  
$\kappa \approx \sqrt{2}$.
 For $\zeta=0.25$  this  
  gives $R\approx 4$, in  agreement with the experimental 
 value  $5.3 \pm 1.8$.

{\bf 8.} In conclusion, we have 
analyzed the factorization breaking observed 
in dijet diffractive production at Tevatron, using the same framework which 
allowed to explain the factorization breaking between HERA and Tevatron single 
diffractive processes. In both cases, the long distance diffractive 
interactions between the proton and the antiproton are the cause of the 
factorization breaking mechanisms.

Interestingly enough, the  pattern of factorization breaking in this framework 
is quite dependent of the number of rapidity gaps (``no-gap'', ``single-gap'' 
and ``two-gap''), in agreement with the experimental findings. The 
factorization breaking  is strong for the comparison between 
``no-gap''and  ``single-gap'' cross-sections, since it is related to the total 
$p\bar p$ absorption factor (\ref{4}), while it is weak\footnote{Note that our 
derivation  gives an interpretation of 
the empirical ``gap  probability renormalization'' proposed in Ref.\cite{dino}. 
In particular, only little price (see {\it e.g.} formula (\ref{ratiosbis})) is 
payed for the formation of a second rapidity gap  once the first is present.} 
for ``single-gap'' {\it 
vs.} ``two-gap'' ones, where it is related  to  the fraction of the inelastic 
diffraction to the total inelastic cross-section which is  small.

Thanks to  the identification of a factorization breaking mechanism, an outcome 
of our approach is the numerical evaluation of the ratios $ R_{i/j}$ and in 
particular $R_{2/1}$ which is found experimentally to be ${\cal O}(1),$ that is 
surprisingly high. 
With quite reasonable values for the absorption parameters we 
find results in nice agreement with the data.  Stimulating dijet production 
results, expected coming soon \cite {mor} from the Tevatron, Run II, will allow 
us to perform a more differential analysis of the mechanism we propose.

\vspace{0.3cm}
{\it Note added}.
While completing this study, the paper Ref. \cite{kai} has 
appeared, treating the same question within the rapidity gap formalism. We note 
an  agreement between the two approaches.
\vspace{0.3cm}

\vspace{0.3cm}
{\bf Acknowledgements}
\vspace{0.3cm}

Discussions with Dino Goulianos are highly appreciated. A.B. thanks the Theory
Department of the Saclay Centre for a kind hospitality. This
investigation was supported in part by the by Subsydium of Foundation
for Polish Science NP 1/99 and by the Polish State Commitee fo
Scientific Research (KBN) Grant No 2 P03 B 09322 (2002-2004).

\eject


\begin{thebibliography} {99}
\bibitem{F} T.  Affolder et al. , CDF Coll.  , {\it Phys.  Rev.  Lett.}  
{\bf
85 } (2000) 5043.
\bibitem{G} For a review of CDF results on Diffraction: K.  Goulianos, {\it 
Nucl. Phys. Proc. Suppl.} {\bf 99A} (2001) 37.
\bibitem{HF} T.  Affolder et al. , CDF Coll.  , {\it Phys.  Rev.  Lett.}  
{\bf
84 } (2000) 4215.
\bibitem{B}
A.Bialas, {\it Acta Phys. Polon.} {\bf B33} (2002) 2635. 
\bibitem{GW}
M.L. Good and W.D. Walker
, {\it Phys.  Rev.}  
{\bf
120} (1960) 1857.
\bibitem{FV}
K.Fialkowski and L.Van Hove, {\it 
Nucl. Phys.} {\bf B107} (1976) 211.
\bibitem{MP}
H.Miettinen and J.Pumplin, {\it Phys.  Rev.}  
{\bf
D18} (1978) 1696.
\bibitem{BP} A.Bialas, R.Peschanski, {\it Phys.  Lett.} {\bf B378} (1996) 302;   
{\bf B387} (1996) 405. 
\bibitem{GL}
R.J. Glauber
, {\it Phys.  Rev.}  
{\bf
100} (1955) 242.
\b{OR}
Data for elastic scattering:\\
Total cross-section: E811 Coll., C.Avila {\it et al., Phys. Lett.}{\bf B445} 
(1999) 419;\\
Elastic slope: N.A. Amos 
{\it et al., Phys. Rev. Lett.}{\bf 63} (1989) 2784.
\b{phot}
 For a recent review:  K. Voss,  {\it Vector meson production at HERA},  
XXXVIIIth Rencontres de Moriond  {\it QCD and Hadronic interactions at high 
energy}, March 23rd-29th 2003, hep-ex/0305052.
\b{pum}
J. Pumplin, {\it Phys.  Rev.}  
{\bf
D8} (1973) 2899.
\b{BPR} M. Boonekamp, 
R. Peschanski, 
C. Royon, 
{\it Phys. Rev. Lett.} {\bf 87} (2001) 251806; hep-ph/0301244, to 
appear in {\it 
Nucl. Phys.} {\bf B}.
\b{KMR}
A.  Bialas and
P.~V.~Landshoff, \PLB   256 (1990) 540. V.~Khoze, A.~Martin, 
M.~Ryskin, \EPC 24 (2002) 459 and references therein. \
\b{dino}
 K. Goulianos, J. Montanha, {\it Phys.Rev.} {\bf  D59} (1999) 114017; 
General review and Refs. in K. Goulianos, {\it Diffraction in QCD}, presented at 
CORFU-2001, Corfu, Greece, hep-ph/0203141. 
\b{mor}
M. Gallinaro (representing the CDF collaboration), {\it QCD Results from the 
CDF Experiment at $\sqrt{s}=1.96$ TeV}, XXXVIIIth Rencontres de Moriond  {\it 
QCD 
and Hadronic interactions at high energy}, March 23rd-29th 2003, hep-ph/030421.
\b{kai}
A.B. Kaidalov, V.A. Khoze, A.D. Martin, M.G. Ryskin,  {\it Phys. Lett.} {\bf  
B559} (2003) 235.
\end{thebibliography}
\end{document}